\documentclass[aps,prl,groupedaddress,twocolumn,amsmath,ambsymb]{revtex4-2}

\usepackage{hyperref} \hypersetup{colorlinks=true, citecolor = blue, linkcolor=teal, filecolor=magenta, urlcolor=cyan}

\usepackage{graphicx}
\usepackage{color,colortbl}
\usepackage{xcolor}
\usepackage{soul}

\newcommand{\corrF}{C_{\rm T1}}
\newcommand{\zetaE}{\zeta_{\rm E}}
\newcommand{\zetaI}{\zeta_{\rm I}}
\newcommand{\gdot}{\dot{\gamma}}
\newcommand{\be}{\begin{equation}}
\newcommand{\ee}{\end{equation}}

\newcommand{\std}{\delta}
\newcommand{\corr}{\xi}

\begin{document}

\title{Fluid-Solid Pattern Formation and Strain Localisation via Shear Banding Transition in Model Biological Tissues}

\author{Aidan J. Nicholas}
\author{Suzanne M. Fielding}
\affiliation{Department of Physics, Durham University, Science Laboratories, South Road, Durham DH1 3LE, United Kingdom.}

\date{\today}

\begin{abstract} The rheological properties of biological tissues  are core  to processes such as cancer metastasis, wound healing and embryo development. The emergence of tissue and organ structures during morphogenesis requires the precise formation of spatial patterns. Dating back to Turing, pattern formation has been suggested to arise in tissues via spontaneous symmetry breaking instabilities in the concentration field of chemical morphogens. Within the vertex model of tissue mechanics, we show that spontaneous symmetry breaking may also arise via a mechanical  instability in the strain field of a deformed tissue, leading to a patterned coexistence of fluid and solid regions, with a strong localisation of the strain into shear bands. The nature of the bands differs between tissues in which internal cell-cell dissipation dominates external drag against a substrate, and vice versa.
\end{abstract}

\maketitle

Biological tissues display a host of rheological (deformation and flow) phenomena that are key to processes such as embryo development, tumour progression and wound healing. They are viscoelastic~\cite{forgacs1998viscoelastic}, behaving as deformable solids on short timescales~\cite{phillips1978embryonic} but also showing power law stress relaxation~\cite{khalilgharibi2019stress} and slowly reshaping via internally active processes that include cell rearrangements, division and  death~\cite{guirao2015unified,etournay2015interplay,wyatt2016question}. In nonlinear deformation they display strain stiffening~\cite{harris2012characterizing}. Tissues can also fracture~\cite{bonfanti2022fracture} as a result of externally imposed stretching~\cite{harris2012characterizing} or an organism's own internally generated motility~\cite{prakash2021motility}.

The emergence of tissue and organ structures during morphogenesis~\cite{goodwin2021mechanics,sutlive2022generation,heisenberg2013forces}, for example in gastrulation and neural tube formation, requires the formation of precise spatial patterns. Turing suggested that patterning  may arise in tissues via spontaneous symmetry breaking instabilities in the spatio-temporal reaction-diffusion dynamics of chemical morphogens~\cite{turing1952chemical,turing1990chemical}.  Beyond purely chemical patterning, considerable attention also concerns the role of mechanics~\cite{howard2011turing,sutlive2022generation,heisenberg2013forces,montell2008morphogenetic,lecuit2007cell,odell1981mechanical,hayward2021tissue,farhadifar2007influence,villeneuve2024mechanical,manna2025dynamic}.
For a tissue to reshape itself requires the concerted motion of groups of cells. Tissues experience external  stresses, as well as  active stresses generated internally by actin-myosin networks and transmitted over long length-scales~\cite{sadeghipour2018shear}. The role of mechanical phase transitions is increasingly discussed~\cite{lenne2022sculpting,mao2024mechanical} in processes as diverse as tissue spreading, folding, elongation and segregation~\cite{rustarazo2026tissue}.

Indeed, tissues show spontaneous solid-fluid jamming-unjamming transitions (and vice versa)~\cite{malinverno2017endocytic,park2015unjamming,garcia2015physics,oswald2017jamming}, often linked to changes in cell shape~\cite{park2015unjamming,atia2018geometric}, and sometimes with a coexistence of fluid and solid regions~\cite{mongera2018fluid}. A stress-induced fluidisation transition has also been observed~\cite{cai2022compressive}.  Imaging techniques that track cell trajectories~\cite{keller2013imaging} show that the strain field within a tissue can be heterogeneous, with localised strains~\cite{etournay2015interplay,prakash2021motility,latorre2018active}. 
The concept of Lagrangian coherent structures~\cite{haller2015lagrangian}, used to study flow patterns in fluid mechanics, has been adapted to analyse attracting and repelling organisers of cell trajectories~\cite{serra2020dynamic}. It is therefore natural to ask how mechanical instabilities in general, and fluid-solid transitions in particular, contribute to pattern formation in tissues.

The mechanics of a tissue is governed not only by elastic stresses, but also by frictional drag stemming from the breakage and reformation of cell adhesions. This is often modeled as viscous dissipation~\cite{tong2022linear,tong2023linear}, assuming tissue reshaping to be slow in comparison. It can arise internally within a tissue, via cell-cell adhesions, and externally via adhesive contacts between a tissue monolayer and a supporting substrate. Reducing internal cell-cell adhesions  reduced the correlation length of cooperative cell motions in Ref.~\cite{czirok2013collective}. Increasing external substrate stiffness increased the correlation length  in Ref.~\cite{vazquez2022effect}. Tissue reshaping can also occur without any supporting substrate, e.g. during gastrulation in early embryo development. Actively driven global flows  emerged spontaneously  in a theoretical model of active tissues with internal cell-cell dissipation in Ref.~\cite{rozman2025vertex}, but were suppressed by external cell-substrate dissipation.

In this Letter, we show that a patterned coexistence of fluid and solid regions can arise via spontaneous symmetry breaking in a model biological tissue subject to an imposed deformation. It does so via a mechanical shear banding instability that occurs as the tissue yields from an initially solid-like state. The strain becomes strongly localised in the high shear band(s), with other parts of the sample behaving elastically in comparison. The bands can persist on long timescales, before healing towards a homogeneously fluidised state. The nature of the bands depends strongly on the form of viscous drag.  With internal cell-cell dissipation, a single shear band forms, of width set by the system size. With external cell-substrate dissipation, multiple smaller bands form. 

We work within the widely studied vertex model of tissue mechanics~\cite{nagai2001dynamic,staple2010mechanics,fletcher2014vertex}. However, shear bands arise commonly when soft solids yield~\cite{divoux2010transient,martin2012transient,dimitriou2014comprehensive,fielding2016triggers,manning2007strain,grenard2014timescales,sentjabrskaja2015creep,cohen2006slip,perge2014time} suggesting that our findings may be independent of model and deformation protocol (unidirectional vs oscillatory strain, imposed strain vs. imposed stress, etc.).

{\it Vertex model ---}  A tissue monolayer of confluently packed cells is modeled as a 2D tiling of $c=1\cdots C$ polygons, with each cell defined by the location of its $n_c=1\cdots N_c$ vertices.  The tissue's elastic energy~\cite{farhadifar2007influence,staple2010mechanics}
\begin{equation}
    E=\frac12\sum_{c=1}^C\left[\kappa_A(A_c-A_{c0})^2+\kappa_P(P_c-P_{c0})^2\right].
    \label{eqn:energy}
\end{equation}
The first term on the RHS models cell volume incompressibility in 3D (considering the monolayer's height) by means of an effective 2D area elasticity, with an energy cost when any cell's area $A_c$ differs from a target $A_{c0}$.  The second term models cell-edge tension,  with an energy cost when any cell's perimeter $P_c$ differs from a target $P_{c0}$, governed by a competition between cell cortical contractility and cell-cell adhesion.  The area and perimeter stiffnesses $\kappa_{\rm A}$ and $\kappa_P$ are assumed the same for all cells in the tissue, as is the target cell shape $p_0=P_{c0}/\sqrt{A_{c0}}$. 

The tissue's dynamics is governed by the interplay  between the  resulting elastic forces, with force  $\Vec{F}_n=-\frac{\delta E}{\delta\Vec{x}_n}$ on the $n$th vertex (as sketched in Fig.~\ref{fig:VMdiagram} in the End Matter), and viscous dissipation, which we prescribe below.  We also allow plasticity~\cite{popovic2021inferring,nguyen2025origin} via $T1$ cell rearrangements: when the edge connecting any two neighbouring vertices shrinks below a small length $l_{\rm c}$, these two original vertices (at the junction of cells $\alpha\beta\gamma$  and  $\alpha\beta\delta$, say) are replaced by new ones at the junctions of cells $\beta\gamma\delta$ and $\alpha\gamma\delta$, with a new small edge length $2l_{\rm c}$, giving cell neighbour exchange. We have checked that no rapid reversible T1 events occur, and for robustness against variations in the (small) value of $l_{\rm c}$.

A key finding in what follows will be that the yielding of biological tissues, as predicted by the vertex model, depends strongly on the type of dissipation a tissue experiences. We study two different forms of dissipation. First, we consider dissipation dominated by drag of the constituent cells against an external substrate. Here the  position $\Vec{x}_n$  of the $n$th vertex moves with velocity:
\be
\vec{v}_n\equiv \frac{d \vec{x}_{n}}{dt} = \zetaE^{-1}\vec{F}_{n},
\label{eqn:external}
\ee
with drag coefficient $\zetaE$. Second, for a tissue monolayer  in which any  external substrate drag is negligible compared with in-plane drag  impeding relative motion of neighbouring vertices internally within the tissue,  we have~\cite{rozman2025vertex}
\be
\sum_{m\in S_n}(\vec{v}_n - \vec{v}_m) = \zetaI^{-1} \vec{F}_n.
\label{eqn:internal}
\ee
Alternatively, we can think of this model with internal drag as a simplified 2D projection of a bulk 3D tissue. The sum in Eqn.~\ref{eqn:internal} runs  over the set $S_n$ of all neighbouring vertices $m$ of any vertex $n$. For the tissue as a whole this gives a linear system $M\cdot V=\zetaI^{-1} F$ in the vectors $V$ and $F$ of vertex velocities and forces. Each row of the sparse matrix $M$ has  $+3$ on the diagonal and three $-1$ entries  at locations determined by the  vertex neighbour connectivities, updated after each T1 event. Applying Eqn.~\ref{eqn:internal} to every vertex  would render $M$ singular, due to Galilean invariance. Choosing a reference frame modifies one line of the linear system, lifting this degeneracy. The system's centre of mass velocity is zero in all the results presented. Relative drag is widely used in soft matter particle simulations, eg.~\cite{tighe2010model,groot1997dissipative}.

{\it Protocol ---} We initialise a square Voronoi tiling of polygons~\cite{voronoi1908nouvelles},
render it biperiodic using the algorithm of Ref.~\cite{yan2011computing}, then evolve the tissue to equilibrium for any target cell shape $p_0$, taking RMS vertex velocity  $< 10^{-8}$ as our criterion for equilibrium.  In equilibrium, the vertex model shows a transition at a critical  $p_0=p_0^*\approx 3.81$~\cite{bi2015density,bi2016motility,moshe2018geometric}.
For  $p_0<p_0^*$ it behaves as a solid with finite linear shear modulus. For $p_0>p_0^*$ it is floppy in linear response, with no resistance to small shear deformations, but stiffens beyond a finite imposed strain, $\gamma>\gamma^*(p_0)$~\cite{huang2022shear}. Snapshot cell tilings are shown in the End Matter.

At time $t=0$, we switch on a shear strain of rate $\gdot$, with flow direction $x$ and shear gradient $y$, and Lees-Edwards periodic boundary conditions~\cite{lees1972computer}. In a tissue with external dissipation, vertex $n$ then moves as
$\Vec{v}_n=\zetaE^{-1}\Vec{F}_n+\gdot y_n \Vec{\hat{x}}$, with a shear term added to Eqn.~\ref{eqn:external}. With internal dissipation, co-sliding of the stacked periodic boxes is implemented via a velocity source vector $\vec{G}$ with entries $\gdot L_y$ (resp. $-\gdot L_y$) for any vertex whose connector to its neighbour crosses the domain's top (resp.  bottom) boundary, giving the  modified system $M\vec{V} = \zetaI \vec{F} + \vec{G}$, counterpart to Eqn.~\ref{eqn:internal}. This shear communicates across the tissue via the drag between neighbouring vertices. 

{\it Parameters ---} We simulate a $50:50$ bidisperse tiling of $C$ cells of  target areas $A_0=1, 1.4$, which sets our unit of length. We have checked that the phenomena we report are robust to the shape of the distribution of cell areas, by also simulating a log normal and a Gaussian distribution.  We use  units of modulus and time in which $\kappa_A=1$ and $\zetaE=1$ (resp. $\zetaI=1$) in the external (resp. internal) model.  We choose $\kappa_p=1.0$. Our numerical timestep $\Delta t = 0.01$ and the threshold for T1 events $l_{\rm c}=0.02$. Parameters to vary will be the target cell shape $p_0$ (the same for all cells, with target cell perimeters set accordingly), imposed  shear rate $\gdot$ and system size $C$. 

\begin{figure}[!t]
    \centering
        \includegraphics[width=\columnwidth]{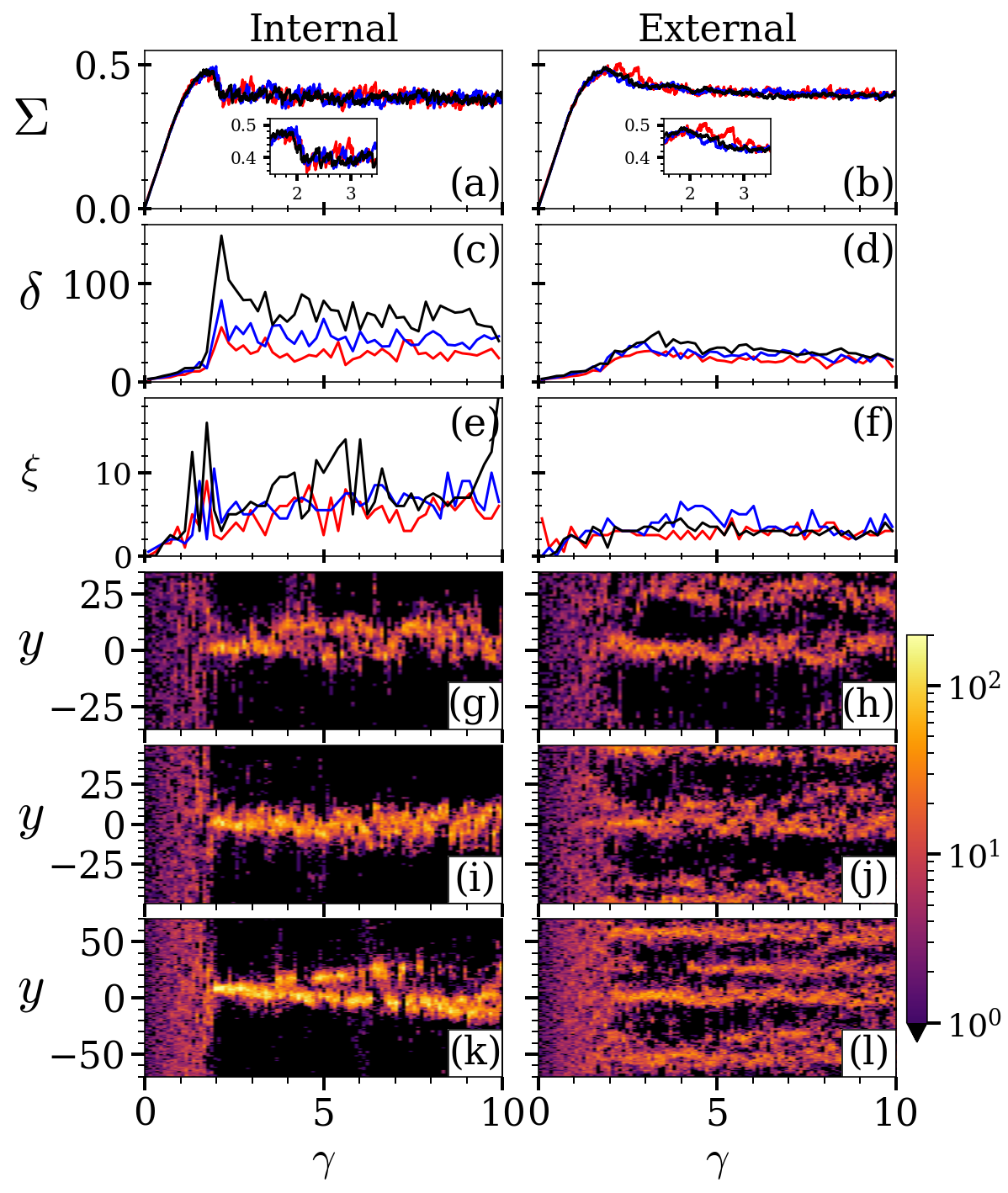}
    \caption{Yielding phenomenology of the vertex model with internal dissipation (left) and  external  dissipation (right). Shown as a function of strain $\gamma$ are the   stress $\Sigma$ {\bf (a+b)},  standard deviation of plasticity  $\std$ {\bf (c+d)}, and   correlation length $\corr$ {\bf (e+f)}, with $\std$ and $\corr$ defined in the main text. System sizes $C = 4096, 8192, 16384$ in red, blue and black curves respectively.     Data in (c-f)  are averaged over strain bins of size $\Delta \gamma=0.2$. Distribution of T1 events across the flow-gradient direction $y$  versus  strain $\gamma$ (in a single realisation, binned over $\Delta\gamma = 0.1$ and $\Delta y= 1$) for  $C = 4096$ {\bf (g+h)}, $C=8192$ {\bf (i+j)}  and  $C=16384$ {\bf (k+l)}.  No T1 events arise in black regions.  $p_{0} = 3.5$,  $\dot{\gamma} = 10^{-4}$.
    }
    \label{fig:pheno}
\end{figure}

{\it Results ---} In exploring yielding as a function of time $t$ or (essentially equivalently) imposed strain $\gamma=\gdot t$ since the inception of shear of constant rate $\gdot$, we shall report the global stress-strain curve $\Sigma(\gamma)$, with $\Sigma$  the $xy$ component of the stress tensor $\Sigma_{ij}(t)=\tfrac{1}{N_c}\sum_{n=1}^{N_c} F_{ni}x_{nj}$, where the sum runs over all vertices in the tiling.  We shall also study the way in which the strain field within the tissue becomes localised into shear bands with layer normals in the gradient direction $y$, and quantify this shear banding using measures to be defined below. 

Figs.~\ref{fig:pheno}(a+b) show the global stress-strain curves for a tissue with internal and external dissipation respectively. In each case, the stress initially rises almost linearly with imposed strain, signifying near elastic response. For $p_0>p_0^*$, the initial response would be floppy, before strain induced solidifcation to an elastic regime beyond a critical $\gamma_{\rm c}(p_0)$~\cite{huang2022shear}. The stress then  attains a maximum, beyond which the tissue yields and the stress decreases to a constant, signifying a state of plastic flow.  With external dissipation, the decrease in stress during yielding happens in a smooth and gradual way. With internal dissipation, the decrease is relatively smooth for small system sizes, but becomes progressively more precipitous for larger systems.

Panels ~\ref{fig:pheno}g-l) show density maps of the number of plastic T1 rearrangement events that arise as a function of imposed strain $\gamma$,  spatially resolved across the flow gradient  direction $y$. At short times (small strains) plastic events are rare, and any that do arise are distributed essentially uniformly across $y$.  A major change then occurs as the tissue yields. The plastic events become much more numerous and furthermore organise into a spatially localised band (or bands), with layer normals along $y$. The $y$-origin is chosen to coincide the $y$-bin in which the most T1 events occur for  $0<\gamma<10$. We have checked that the distribution of the locations of this $y$-bin, before shifting to $y=0$, is uniform and with no bias resulting from the periodic boundary. In no case did we observe localisation along $x$. Because each plastic event leads to an additional forward plastic strain, beyond the globally  imposed one, any band of high plasticity is also one of strong shear. The shear rate integrated across $y$ must however remain equal to the globally imposed one. Accordingly, outwith the plastic bands the shear rate is lower than the imposed rate. Here the tissue deforms essentially elastically, with no plastic events. 

\begin{figure}[!t]
    \centering
        \includegraphics[width=0.95\columnwidth]{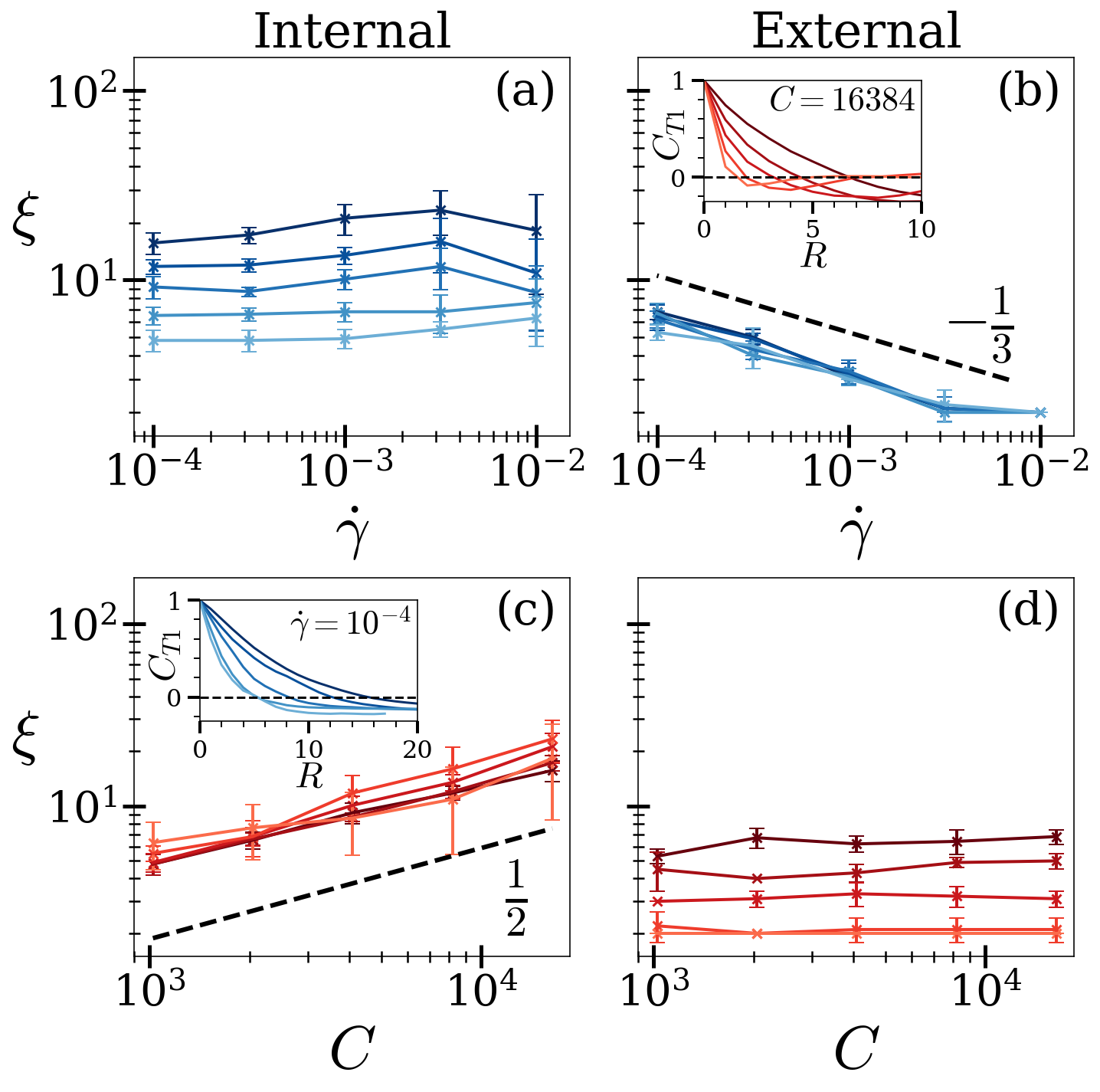}
    \caption{Correlation length in a model with  internal  {\bf (a+c)} and  external  {\bf (b+d)} dissipation, as a function of shear rate for system sizes $C=1024, 2048, .. 16384$ increasing in curves from light to dark {\bf(a+b)}, and as a function of  system size for shear rates $\gdot=10^{-n}$ with $n=0.0, -0.5, \cdots -4.0$, the shear rate decreasing in curves from light to dark {\bf (c+d)}.  Data averaged over the strain interval $2.0<\gamma<6.0$ and over $10$ realisations.  Dashed lines show power laws indicated, as a guide to the eye. Shape parameter $p_0 = 3.5$. Insets show the correlation functions used to extract the lengths.
     \label{fig:corrAll}
     }
\end{figure}

Comparing the left and right columns of Fig.~\ref{fig:pheno} reveals a qualitative difference in the nature of the bands in tissues with internal versus external dissipation. With internal dissipation, a single macroscopic band develops, of a size that increases with system size in g) to k) downwards. With external dissipation, multiple narrow bands develop, which increase in number at roughly fixed band width as the system size increases in h) to l) downwards.

We now define two measures to quantify the banding, as evolving functions of the imposed strain $\gamma$. During each strain increment $\gamma\to\gamma + \Delta \gamma$, with $\Delta\gamma=0.1$, we divide the  $y$ axis into bins of size $\Delta y=1.0$, and count the number of T1 events that occur in each bin. The standard deviation $\delta(\gamma)$ in this number across $y$ then measures  how strongly the plasticity (and so strain rate) is banded across $y$. The extent $\corrF(\Delta y)$ to which the  number of plastic events at any $y$ is correlated with that at $y+\Delta y$, in any strain increment $\gamma\to\gamma + \Delta \gamma$, further defines a correlation length $\xi(\gamma)$, via the value of $\Delta y$ at which this correlation function $\corrF(\Delta y)$ falls to zero. Accordingly,  the evolving standard deviation $\delta(\gamma)$ measures the evolving {\em amplitude} of shear banding and the correlation length $\xi(\gamma)$ the characteristic {\em width} of the bands. 

\begin{figure}[!t]
    \centering
        \includegraphics[width=0.95\columnwidth]{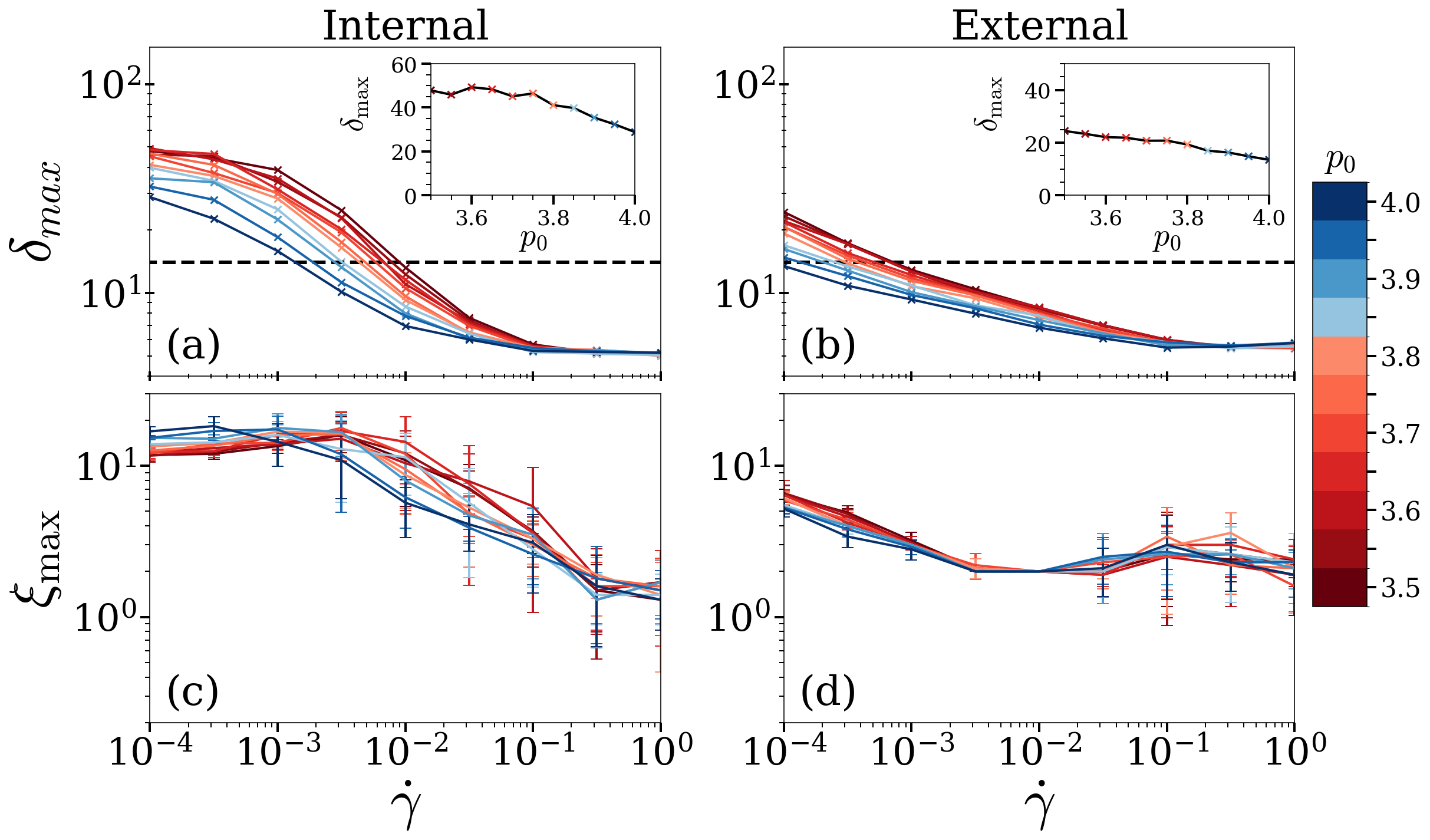}
    \caption{Amplitude {\bf (a+b)} and correlation length {\bf (c+d)} of shear banding in a model with  internal dissipation  {\bf (a+c)} and  external dissipation {\bf (b+d)},
    plotted as a function of shear rate for values of the shape parameter $p_0=3.50, 3.55, \cdots 4.00$
    with a diverging colour map centred on $p_0^*=3.81$. Data averaged over $10$ realisations. System size $C=8192$. In snapshots of the kind shown in Fig.~\ref{fig:pheno}g-l) and ~\ref{fig:longTimes}c-f), shear bands are visually apparent for values of $\delta_{\rm max}$ above the dashed line. 
}
    \label{fig:shape}
\end{figure}

At early times (small strains), the standard deviation $\delta(\gamma)$ is small (Figs.~\ref{fig:pheno}c+d), consistent with near uniform shear. It then rises dramatically as shear bands form. With external dissipation, the standard deviation and correlation length (panels d+f) are almost  independent of system size, consistent with a spatial pattern of bands that remains statistically comparable in any macroscopic slice of the system of a given height in $y$, as the system size increases in panels h) to l) downwards. In contrast, with internal dissipation both measures increase with increasing system size (panels c+e), consistent with the formation of a single band that is broader in larger systems from g) to k) downwards. 

Fig.~\ref{fig:corrAll} explores this dependence in more detail. With internal dissipation, the correlation length grows with system size as $\xi\sim C^{m}$ with $m=0.51\pm 0.06$ (panel c), suggesting that  $\xi\sim C^{1/2}\sim L_y$ where $L_y\propto \sqrt{C}$ is the height of our square simulation box, consistent with a single band of width that grows with system size. Recall  Figs.~\ref{fig:pheno}g) to k) downwards. $\xi$ is however relatively independent of shear rate, in this case of internal dissipation (panel a). With  external dissipation the correlation length is relatively independent of system size (panel d), consistent with the formation of an increasing number of bands of roughly constant band width as system size increases in  Figs.~\ref{fig:pheno} h) to l) downwards. The band width however increases with decreasing shear rate, $\xi\sim\gdot^{-p}$ with $p\approx 0.34\pm 0.05$ (panel b).

So far, we have considered a fixed value of the target cell shape  in the regime for which the tissue is solid in linear response, $p_0=3.5<p_0^*$. The phenomenology we report is however qualitatively robust to changing $p_0$. To evidence this, we quantify the degree of shear banding  across a full range of $p_0$ (and shear rate $\gdot$)  via the standard deviation $\delta(\gamma)$ and correlation length $\xi(\gamma)$ defined above. For any $p_0$ and $\gdot$, we further maximise these over the strain interval $1.5<\gamma<4.0$ during which shear bands form, when $\delta$ indeed peaks as seen in Fig.~\ref{fig:pheno}c+d). These maximised values $\delta_{\rm max}$ and $\xi_{\rm max}$ are reported in Fig.~\ref{fig:shape} as a function of shear rate for a range of $p_0$ values, below and above $p_0^*$. In each case, $\delta_{\rm max}$ and $\xi_{\rm max}$ increase with decreasing imposed shear rate. As noted above, for $p_0>p_0^*$ the tissue is initially floppy in linear response but  solidifies at modest imposed strains~\cite{huang2022shear}. It then subsequently yields, and shear bands as it does so.

\begin{figure}[!t]
      \includegraphics[width=0.95\columnwidth]{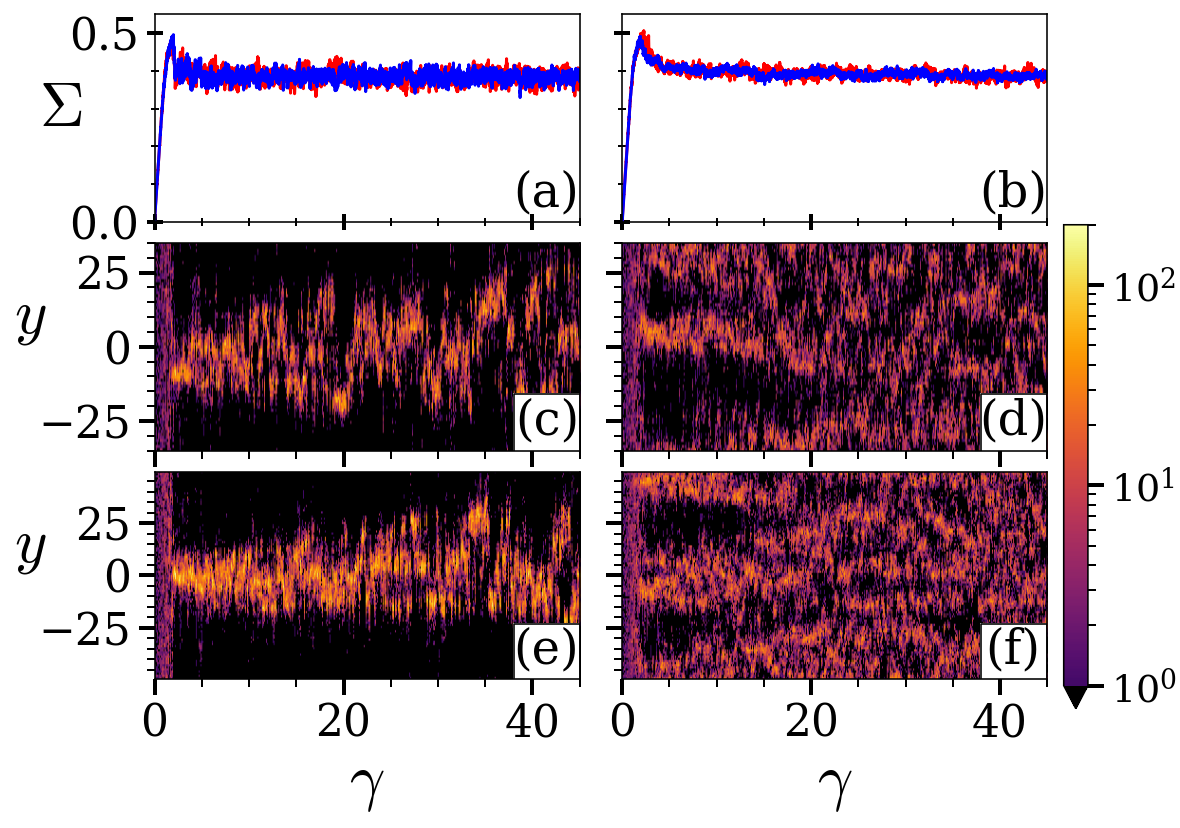}
    \caption{Curves as in Fig.~\ref{fig:pheno} (a+b) and (g-j), but now over the larger strain interval  $0<\gamma<45$ to show the evolution of the shear bands at long times and large strains.}
    \label{fig:longTimes}
\end{figure}

We have shown, then, that the vertex model predicts the formation of shear bands as a tissue first yields  from an initially solid state towards a  plastically flowing one.
Finally in Fig.~\ref{fig:longTimes} we explore the fate of the bands over longer timescales, as shearing continues  to large strains. With internal dissipation (left panels), for which just a single band forms as the tissue first yields, the band progressively broadens over time. This suggests that the tissue may eventually become homogeneously fluidised at very long times, which are however inaccessible computationally, and indeed perhaps unlikely to arise in tissues. In a tissue with external dissipation, for which multiple smaller bands form during yielding, the pattern of banding appears to remain statistically relatively constant over the full duration accessible computationally. Whether the flow finally homogenises on even longer timescales remains an open question. 

{\it Discussion ---} A key finding of our work is that the nature of the shear bands depends on the form of dissipation present. Indeed, an  ongoing debate~\cite{lin2026viscous,rozman2025vertex,karnat2025noninvasive,nguyen2026cell} concerns whether external dissipation against (e.g.) the extracellular matrix dominates  internal dissipation within a tissue, or vice versa. For simplicity, we model the former via the velocity of cell vertices relative to the substrate; in practice, it may arise anywhere on the basal cell sides and/or via active retrograde flows of the actomyosin network~\cite{alert2020physical}. Likewise, we model the latter via relative velocities between neighbouring vertices; in practice, bulk intracellular dissipation may also contribute~\cite{lin2026viscous,karnat2025noninvasive}.

However, the detailed mechanism of dissipation may be less important than its overall form (internal vs. external). Indeed, at the level of a continuum hydrodynamic  description, the distinction arises via the force balance condition, $0=\nabla\cdot \Sigma-\zetaE\vec{v}+\zeta_{\rm I}\nabla^2\vec{v}$, the shear component of which in Fourier space ($\partial_y \to ik$)  gives $0=-k^2\delta\Sigma_k - \zetaE\delta\gdot_k - \zeta_{\rm I}k^2\delta\gdot_k$, for small precursor heterogeneities $\delta\Sigma_k,\delta\gdot_k$ in the elastoplastic shear stress and shear rate about  a state of initially homogeneous shear.

A constitutive model can then be used to express $\Sigma=\Sigma(p,\sigma,a)$ in terms of the cell perimeter $p$, cell orientation $\sigma$ and rate of plastic events $a$, with  dynamical equations $\partial_t p=f(\gdot,p,\sigma,a)$, $\partial_t \sigma=g(\gdot,p,\sigma,a)$, $\partial_t a=h(\gdot,p,\sigma,a)$, with spatially local functions $f,g,h$~\cite{fielding2023constitutive}. Linearising and eliminating $\delta\gdot_k$ via force balance gives $\omega_k (\delta \sigma,\delta p, \delta a)_k^T=M_k\cdot (\delta \sigma,\delta p, \delta a)_k^T$, with the growth rate $\omega_k$ of small heterogeneities at wavevector $k$ determined by the matrix $M_k$. With internal dissipation only ($\zeta_{\rm E}=0$), the $k-$dependence divides out of the force balance condition and $M$ is independent of $k$. Stabilising diffusive terms arising via a breakdown of locality at the cellular level then map $\omega_k\to \omega_k-Dk^2$ and  bands grow most quickly as $k\to 0$, i.e., at the box size~\cite{fielding2003early,lu2000effects,spenley1996nonmonotonic}, as seen in our simulations of the vertex model. With external dissipation only ($\zetaI=0$), $M$ also contains unstable terms in $k^2$ and bands develop more quickly on smaller lengthscales, as also seen in our vertex model simulations. This calculation also naturally incorporates mixed dissipation, which will form the focus of future work.

In summary, we have predicted that a mechanical shear banding instability arises in a model biological tissue during yielding from an initially solid-like state towards a plastically flowing state. This leads to a patterned coexistence of solid and fluidised regions, with a localisation of the imposed strain in the high shear band, the rest of the tissue behaving essentially elastically in comparison. It remains an open challenge to elucidate the fate of the bands at long times. It would be interesting in future to study whether shear bands might also form in tissues during yielding under imposed stress~\cite{grenard2014timescales,sentjabrskaja2015creep} and oscillatory shear~\cite{cohen2006slip,perge2014time}.

{\it Acknowledgements ---}  This project has received funding from the European Research Council (ERC) under the European Union's Horizon 2020 research and innovation programme (grant agreement No. 885146).


%

\section*{End Matter}

\begin{figure}[!t]
      \includegraphics[width=0.95\columnwidth]{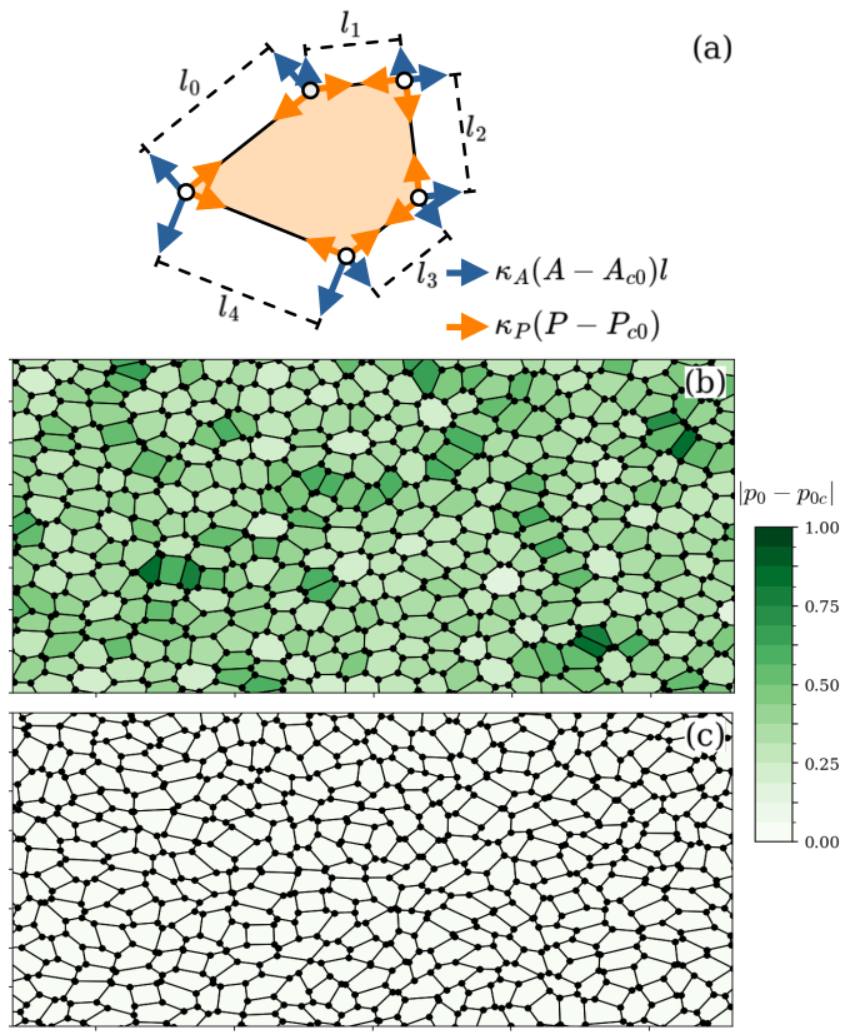}
    \caption{(a) Sketch of a cell in the vertex model showing the forces that act.  Snapshot cell tiling in a tissue with (b) $p_0=3.5$, in the solid phase and (c) $p_0=4.0$, in the fluid phase. The colour scale shows the difference between each cell's actual shape and its target shape.}
    \label{fig:VMdiagram}
\end{figure}

In Fig.~\ref{fig:VMdiagram}, we show a sketch of the forces that act at the cell level in the vertex model. We also show representative snapshot cell tilings for a tissue in the solid and fluid phase of the model.

\end{document}